\documentclass{article}
\usepackage{amsthm}
\newtheorem{theorem}{Theorem}
\newtheorem{lemma}[theorem]{Lemma}
\newtheorem*{problem}{Problem}%
\newtheorem*{definition}{Definition}

\newcommand{\institute}[1]{\date{#1}}
\newcommand{\email}[1]{ {\small \tt{#1} } }

\usepackage{graphicx,amssymb,amsmath}

\usepackage{subfigure}

\newcommand{\comment}[1]{}

\graphicspath{ {.}, {figures/} }

\newcounter{stepc}
\newcounter{substepc}
\newenvironment{astep}[1][0]{
\addtocounter{stepc}{#1}
\par
\bigskip
\noindent\ignorespaces %
\textbf{Step \arabic{stepc}.} \itshape
}{
\par
\ignorespacesafterend
\addtocounter{stepc}{1}
\bigskip
}
\newenvironment{asubstep}{
\par
\bigskip
\noindent\ignorespaces %
\textbf{Step \arabic{stepc}.\arabic{substepc}.} \itshape
}{
\par
\ignorespacesafterend
\addtocounter{substepc}{1}
\medskip
}

\title{Computation of the Hausdorff distance between sets of line segments in parallel\thanks{This research was performed in scope of the DFG-project ``Parallel algorithms in computational geometry with focus on shape matching'' under the contract number AL 253/7-1.}}

\author{Helmut Alt \and Ludmila Scharf}
\institute{Freie Universit\"at Berlin,\\
\email{\{alt,scharf\}@mi.fu-berlin.de}}

\index{Alt, Helmut}
\index{Scharf, Ludmila}

\begin{document}
\thispagestyle{empty}
\maketitle

\begin{abstract}
   We show that the Hausdorff distance for two sets of non-intersecting line segments can be computed in parallel in $O(\log^2 n)$ time using $O(n)$ processors in a CREW-PRAM computation model. 
   We discuss how some parts of the sequential algorithm can be performed in parallel using previously known parallel algorithms; and identify the so-far unsolved part of the problem for the parallel computation, which is the following:  
   Given two  sets of $x$-monotone curve 
   segments, red and blue,  for each red segment find its extremal intersection points with the blue set, i.e. points with the minimal and maximal $x$-coordinate. Each segment set is assumed to be intersection free. 
   For this intersection problem we describe a parallel algorithm which completes the Hausdorff distance computation within the stated time and processor bounds.
\end{abstract}
\section{Introduction}
\label{sec:intro}
Evaluating the similarity of two geometric shapes is an important problem in different fields of computer science including computer vision and pattern recognition. One of the most natural similarity measures is the \emph{Hausdorff distance} which is defined for any two compact sets. The directed Hausdorff distance between two compact point sets $P$ and $Q$ is defined as $d_H(P,Q)=\max_{p\in P} \min_{q\in Q}d(p,q)$, where $d(p,q)$ denotes the Euclidean distance between the points $p$ and $q$. The (undirected) Hausdorff distance $D_H(P,Q)$ is defined as maximum of the two directed distances: $D_H(P,Q)=\max\left\{ d_H(P,Q), d_H(Q,P) \right\}$. If we can compute the directed Hausdorff distance, we can clearly compute the undirected distance within the same asymptotic time bounds.
%
\comment{
In this paper we describe a parallel algorithm for the following intersection problem between two sets of  curve segments:
\begin{problem}\label{problem}
   Given two \emph{well-behaved} sets $A$ and $B$ of curve segments in the plane,  for each  segment $a\in A$ find the intersection point of $a$ with the set $B$ with the smallest $x$-coordinate (if exists), where well-behaved means that all curve segments are $x$-monotone, the intersection between any two segments of the same set is empty except possibly at endpoints, and all intersections between any two segments can be computed in constant time. 
\end{problem}

This rather specific variant of the intersection problem arises as a part of the Hausdorff distance computation for the sets of line segments: The directed Hausdorff distance between two, not necessarily discrete, point sets $P$ and $Q$ is defined as $d_H(P,Q)=\sup_{p\in P} \inf_{q\in Q}d(p,q)$, where $d(p,q)$ denotes the Euclidean distance between the points $p$ and $q$. The (undirected) Hausdorff distance $D_H(P,Q)$ is defined as maximum of the two directed distances: $D_H(P,Q)=\max\left\{ d_H(P,Q), d_H(Q,P) \right\}$. If we can compute the directed Hausdorff distance, we can clearly compute the undirected distance within the same asymptotic time bounds. 
}
Efficient sequential algorithms are known for the Hausdorff distance computation for $P$ and $Q$ being discrete point sets, sets of non-intersecting straight line segments \cite{alt:behrends} and for algebraic parameterized curve segments \cite{as-chdco-08}. 

In this paper we show that the Hausdorff distance for two sets of non-intersecting line segments can be computed efficiently in parallel within $O(\log^2n)$ time using $O(n)$ processors in the CREW-PRAM computation model. 

Due to the current trend in hardware development, where the performance increase is achieved through additional computing units (processor kernels) instead of increase in CPU speed,  parallel algorithms have gained new popularity in the algorithm development. Some graphic cards support general computations on  graphics hardware (GPGPU) which comprises up to several hundreds of parallel processing units, which even more motivates for development of parallel algorithms. 

In this paper we use the PRAM as model of computation model instead of choosing one of the currently available hardware platforms, since we want to examine principal parallelism of the problem instead of concentrating on specific technical details of some hardware. 
We plan an implementation on a GPGPU platform. Certainly, it is not feasible to implement the PRAM-algorithm presented here directly, but its major underlying ideas should be useful.

We briefly recapitulate here the key steps of the sequential algorithm for computing the directed Hausdorff distance between two sets $P$ and $Q$ of non-intersecting line segments from \cite{alt:behrends}: (1)~Construct the Voronoi diagram $VD(Q)$ of the set $Q$; (2) For each endpoint $p$ of a segment in $P$ find its closest segment in $Q$ using $VD(Q)$ and compute the distance from $p$ to that segment; (3) Determine the so-called ``critical points'' on the edges of $VD(Q)$; (4) For each critical point $q$ compute the distance from $q$ to its nearest segment in $Q$; (5) Return the maximal distance of endpoints and  critical points.

\begin{figure}[htbp]
   \hfill \subfigure[Trademark images represented by sets of line segments]{\includegraphics[page=2,scale=.9]{hd} \includegraphics[page=1,scale=.9]{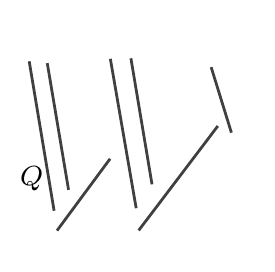}} \hfill
   \subfigure[Critical points]{\includegraphics[page=4,scale=1]{hd}}  \hfill 
   \subfigure[Hausdorff distance]{\includegraphics[page=3,scale=1]{hd}}  \hfill \ 
   \caption{ An example of two sets of line segments. The Hausdorff distance is attained at a critical point on a Voronoi edge.}
   \label{fig:hd}
\end{figure}

The authors show that the critical points, i.e., the points where the directed Hausdorff distance $d_H(P,Q)$ can be attained, besides the endpoints of the segments in $P$, are the intersection points of $P$ with the Voronoi edges of $Q$. Furthermore, they prove that for each edge of $VD(Q)$ only the extreme intersection points, i.e., the first and the last intersection point along the curve segment, 
are  critical points (s.~Figure~\ref{fig:hd}), thus reducing the total number of critical points to $O(n)$.
For $x$-monotone curves the extreme intersection points are the points with the minimal and maximal $x$-coordinate. A non-$x$-monotone parabola segment can be split into two $x$-monotone segments at the point with the vertical tangent. Thus, in the following we can assume that all Voronoi edges are $x$-monotone. 


For all but one steps of the above algorithm there is either a straightforward parallel version or, due to some previous work, it is known how to compute them in parallel:
A parallel algorithm for computing the Voronoi diagram of a set of line segments is given in~\cite{DBLP:journals/algorithmica/GoodrichOY93}. It runs in $O(\log^2 n)$ time on a $O(n)$ processor CREW-PRAM and uses the divide-and-conquer technique.
For a given planar subdivision with $n$ vertices a point location data structure supporting $O(\log n)$-time queries can be constructed in $O(\log n)$ time on an EREW-PRAM with $O(n)$ processors~\cite{Tamassia:1989:OPA:72935.72978}. Thus, step (2) can be performed efficiently in parallel.
Parallelization of  steps (4) and (5) is straightforward, since the distance for each of the $O(n)$ critical points to the nearest segment in $Q$ can be computed independently. 

The only open question in the parallel computation of the Hausdorff distance is to determine the critical points, i.e., intersection points between segments and Voronoi edges, efficiently.
The sequential algorithm uses the plane sweep technique, in which the endpoints of the  line and parabola segments are sorted by $x$-coordinate and a vertical line is swept from left to right across the scene. Each time an intersection between an edge $e$ of $VD(Q)$ and a segment in $P$ is detected the edge $e$ is removed from scene, thus preventing  computation of further intersection points of the same edge. Another sweep is preformed from right to left to detect the intersection points with the maximal $x$-coordinate.

Clearly, the sweepline technique is inherently sequential, and we therefore need different tools to compute the critical points in parallel. For the related intersection detection problem: Given $n$ line segments in the plane, determine if any two of them intersect --  there exist efficient parallel algorithms. 
In~\cite{Aggarwal:1988sf}  a CREW-PRAM algorithm is presented that runs in $O(\log n)$ time and uses $O(n\log n)$ processors. This result is improved in \cite{68482}, where intersection detection is performed in $O(\log n)$ time on a CREW-PRAM with $O(n)$ processors. The latter algorithm has optimal processor-time product because  of the $\Omega(n\log n)$ sequential lower bound for this problem \cite{Preparata:1985:CGI:4333}. 

For the intersection reporting problem, i.e., finding all pairwise intersections between line segments, a CREW-PRAM algorithm with running time $O(\log^2 n)$ on $O(n+k/\log n)$ processors is given in \cite{Goodrich:1989:ILS:72935.72950}, where $k$ is the number of reported intersections. Detecting all intersections between two sets of line segments, each of which is intersection free, can be performed in $O(\log n)$ time using $O(n+k/\log n)$ processors in a CREW-PRAM model \cite{Goodrich:1990:PMV:98524.98539}.

\newcommand{\pname}{first-last intersection problem}
We call the intersection problem arising in the Hausdorff distance computation, which also may be of independent interest, the \emph{\pname} (defined below). Since edges of the Voronoi diagram of a set of line segments can be line or parabola segments, we want our algorithm for this intersection problem to work for more general sets of curve segments: 
\begin{definition}
   Two sets of curve segments $A$ and $B$ are called \emph{well-behaved} if every segment in $A\cup B$ is $x$-monotone; no two segments of the same set have a common point except possibly common endpoints; any two segments from different sets intersect at most twice; all intersections between any two segments can be computed in constant time; and for every segment we can compute in constant time for a given $x$-coordinate the corresponding $y$-coordinate.
\end{definition}

Observe that if we split the parabola segments of the Voronoi diagram at the points with vertical tangent, then the set $P$ of line segments and the Voronoi edges of $Q$ are two well-behaved sets.  The problem of finding critical points for the Hausdorff distance can then be formulated as:
\begin{problem}[First-Last Intersection Problem]
  Given two well-behaved sets $A$ and $B$ of curve segments in the plane,  for each  segment $a\in A$ find the intersection points of $a$ with the segments from the set $B$ with the smallest and with the largest $x$-coordinate. 
\end{problem}

All above mentioned line segment intersection parallel algorithms utilize the \emph{segment tree} data structure \cite{Bentley:1980:OWC:1311064.1311126}, which is also used in this paper and is described in section~\ref{sec:segT}. There is no trivial modification of the  mentioned algorithms for intersection detection or intersection reporting problems which yields an efficient algorithm for \pname. 
Here we present an algorithm that solves this problem in $O(\log^2 n)$ time on a $O(n)$ processor CREW-PRAM and thus prove the following theorem:
\begin{theorem}
   Let  $A$ and $B$ be two well-behaved sets of curve segments in the plane with $\vert A\vert +\vert B\vert =n$. The \pname\ for $A$ and $B$ can be solved in $O(\log n)$ time on $O(n \log n)$ processors using $O(n\log n)$ storage in the CREW-PRAM model. Alternatively, the problem can be solved in $O(\log^2 n)$ time on $O(n)$ processors.
   \label{la_2}
\end{theorem}

Theorem \ref{la_2} together with the above mentioned previous work completes the proof of Theorem~\ref{thm}:
\begin{theorem}
   Given two sets $P$ and $Q$ of $n$ line segments, such that no two segments of the same set intersect, except possibly at the endpoints, the Hausdorff distance $D_H(P,Q)$ can be computed in $O(\log^2 n)$ time on $O(n)$ processors using $O(n\log n)$ storage  in the CREW-PRAM model.
   \label{thm}
\end{theorem}

\section{Segment Tree and Interval Tree}
\label{sec:segT}
In this section we briefly describe two data structures used by our algorithm. 

\paragraph{Segment Tree.} Let $S$ be a set of $n$ line segments. Sorting the $2n$ endpoints of the line segments  by $x$-coordinate and projecting them onto the $x$-axis results in at most $2n+1$ intervals. A \emph{segment tree} for the set $S$ is a complete balanced binary tree $T$ with the following properties: Each of the $2n+1$ intervals is stored at a leaf of $T$ in sorted order. An internal node $v$ of $T$ stores the interval that is the union of the intervals of its children.

\begin{figure}[htbp] 
   \subfigure[A set of four segments and the corresponding segment tree $T$ with cover-lists of the nodes of $T$.]{ \qquad \label{fig:segT} \includegraphics[scale=.8]{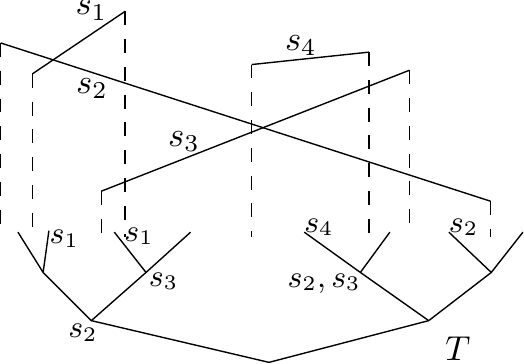}\qquad} \hfill
   \subfigure[A set of seven intervals and the corresponding interval tree with unsorted interval lists at the nodes.]{\qquad\label{fig:intT}\includegraphics[scale=.8]{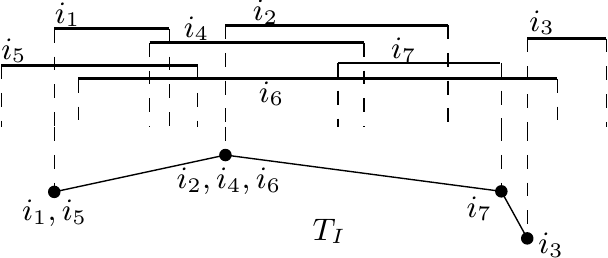}\qquad}
   \caption{Examples of a segment and an interval tree.}
\end{figure}

A segment $s\in S$ \emph{covers} a node $v$ if the interval of $v$ is completely contained in the projection of $s$ onto the $x$-axis, but the parent interval of $v$ is not. Every node $v$ of $T$ stores a \emph{cover-list} -- the list of segments that cover $v$. 
Additionally, every node $v$ stores a list of segments that have at least one endpoint in the interval of $v$ -- the \emph{end-list}. An example of a segment tree with cover-lists of the nodes is given in Figure~\ref{fig:segT}.

Since every segment is contained in the cover-lists of at most two nodes of each level of $T$ and in at most two end-lists per level, the size of $T$ is $O(n\log n)$. A segment tree with cover- and end-lists can be constructed in $O(\log n)$ time on a $O(n)$ processor CREW-PRAM (see e.g.~\cite{Aggarwal:1988sf}).

\paragraph{Interval Tree.} Let $I$ be a set of $n$ closed intervals on the real line. An \emph{interval tree} $T_I$ that stores $I$ is defined recursively as follows: If $I=\emptyset$, $T_I$ is a leaf. Otherwise, the root node $v$ of $T_I$ stores a reference value $r_v$ and a list of the intervals of $I$ that contain $r_v$. The left (right) child of $v$ is an interval tree for the intervals in $I$, whose right (left) endpoint is strictly less (greater) than $r_v$. 
Typically, the reference value $r_v$ is chosen to be the median of the endpoints in $I$. This ensures that the height of $T_I$ is $O(\log n)$. The intervals of a node of $T_I$ are stored twice: in one list sorted by the first endpoint, and in the second, sorted by the second endpoint (see e.g.~\cite{4M} for a detailed description). An example of an interval tree is given in Figure~\ref{fig:intT}.
\comment{
\begin{figure}[htbp]
   \begin{center}
      \includegraphics[scale=.8]{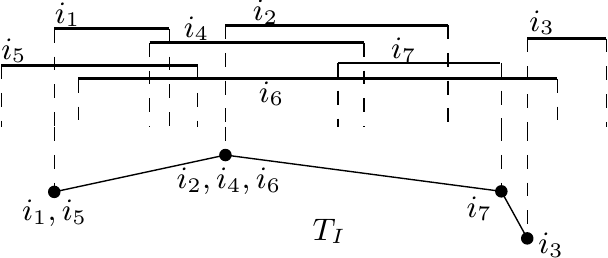}
   \end{center}
   \caption{A set of seven intervals and the corresponding interval tree with unsorted interval lists at the nodes.}
   \label{fig:intT}
\end{figure}
}

An interval tree for a set of $n$ intervals uses $O(n)$ storage and can be build in $O(n\log n)$ time with a sequential algorithm. Using the interval tree we can report all intervals that contain a query point in $O(\log n +k)$ time, where $k$ is the number of reported intervals.

For the parallel construction we can sort the $2n$ endpoints of the intervals and build a complete balanced binary search tree $T_I$ on the values of the endpoints. Now, using the values in the nodes of the search tree as the reference values, each interval of $I$ can independently find the highest node in $T_I$ whose reference value it contains and assign itself to that node. Finally, we  sort the node entries  lexicographically twice: by $(v,i_1)$ and by $(v,i_2)$, where $v$ is an identifier of a tree node and $i_1,i_2$ are the endpoints of an interval assigned to $v$. The intervals can then be written to the two lists of their corresponding nodes: in one sorted by start- and the other sorted by endpoint.  This  gives us a tree with the properties of the interval tree and $O(\log n)$ height. 
Thus, using a fast sorting algorithm we have:
\begin{lemma}
   For a given set of $n$ intervals an interval tree can be constructed  in $O(\log n)$ time on $O(n)$ processors in the CREW-PRAM model.
   \label{la:intTree}
\end{lemma}

\section{A Parallel Algorithm for the First-Last-Inter\-section Problem }
\label{sec:algo}

Let $A$ and $B$ be two well-behaved sets (red and blue resp.) of curve segments in the plane with $\vert A \vert + \vert B \vert =n$. 
Here we describe how to find for each segment $a\in A$ the intersection point with $B$ with the minimal $x$-coordinate, the intersections with the maximal $x$-coordinate can be determined symmetrically.

Our algorithm begins with the construction of a segment tree $T$ for the set $A\cup B$:

\begin{astep} 
   Build a segment tree $T$ for $A\cup B$. For each node $v$ of $T$ construct separate cover-lists $C_A(v),C_B(v)$ for sets $A$ and $B$ respectively, and separate end-lists $E_A(v),E_B(v)$. 
   Sort $C_A(v)$ and $C_B(v)$ by $y$-coordinate for all nodes $v$ in parallel.
\end{astep}
\comment{
\begin{step} 
   Build a segment tree $T$ for $A\cup B$. For each node $v$ of $T$ construct separate cover-lists $C_A(v),C_B(v)$ for sets $A$ and $B$ respectively, and separate end-lists $E_A(v),E_B(v)$. 
   Sort $C_A(v)$ and $C_B(v)$ by $y$-coordinate for all nodes $v$ in parallel.
\end{step}
}

Since the initial sets are intersection free, the $y$-order within the cover-lists of each color is well-defined.

Chazelle showed in \cite{808674} that if two line segments of a set intersect, then in the corresponding segment tree there must be a node $v$ such that either both segments are in $C(v)$ or one is in $C(v)$ and the other in $E(v)$. In our example in Figure~\ref{fig:segT} the intersection of the segments $s_2$ and $s_3$ is of the first type and the intersection between $s_1$ and $s_2$ of the second. It is easy to see that a similar statement holds for well-behaved curves as defined in Section \ref{sec:intro}:
Consider an intersection point $p$ of curve segments $s$ and $s'$ and the leaf $v$ of the segment tree that contains the $x$-coordinate of $p$. Then each of the two segments $s$ and $s'$ appears in the cover-list of one node on the path from $v$ to the root of the segment tree. Let $v'$ denote the highest, i.e., closest to the root, such node, and let w.l.o.g.\ $s'$ be in the cover-list of $v'$. Then, either $s$ is also in the cover-list of $v'$, or, since $s$ covers a sub-interval (the interval of $v$) of $v'$ but not the complete interval of $v'$ or any of its ancestors, $s$ must be in the end-list of $v'$. The same argument holds for every intersection point of $s$ and $s'$ and the corresponding leaf of the segment tree. 

In the two-set setting this means that if a red segment $a$ and a blue segment $b$ intersect, then there must be a node $v$ in $T$ such that either (1) $a\in C_A(v)$ and $b\in C_B(v)$, (2)~$a\in E_A(v)$ and $b\in C_B(v)$, or (3) $a\in C_A(v)$ and $b\in E_B(v)$. 
The following steps of the algorithm deal with each of these three cases. Whereas the handling of the first two cases (Step 2) is a  modification of the corresponding steps of the algorithm in \cite{Goodrich:1990:PMV:98524.98539}, the third case demands additional processing (Step~3).

\begin{astep} 
   For each node $v$ of $T$ and each segment $a\in C_A(v)\cup E_A(v)$  do in parallel: Find the  neighbors of $a$ in $C_B(v)$ with respect to the $y$-order at the $x$-coordinate of the leftmost point of $a$ within the interval of~$v$. Compute the intersections of $a$ with its  neighbors, if any exist, and record the one with the minimum $x$-coordinate.
\end{astep}

Since we want to find the intersection point of $a$ with the minimum $x$-coordinate and all curve segments are $x$-monotone, we do not need to find all intersections of $a$ within the interval of $v$ but only those with the  neighbors at the left border of the interval, or at the $x$-coordinate of the left endpoint of $a$, respectively. 

For type (3) of intersections we observe that for a given segment $b\in E_B(v)$ we can easily determine the lowest, $a_1$, and the highest, $a_2$, red segment in $C_A(v)$ intersected by $b$, such that the intersection points have $x$-coordinates inside the interval of $v$, using binary search on $C_A(v)$. Clearly, all red segments in $C_A(v)$ between $a_1$ and $a_2$ are also intersected by $b$ forming a set of consecutive ranks in $C_A(v)$ with respect to its ascending order in $y$-direction. This set we call the \emph{rank interval} of $b$, which has a constant size representation by $a_1,a_2$.

In the following we are going to find for each node $v$ of $T$ the set $I(v)$ of the rank intervals for all $b\in E_B(v)$. Then we process and narrow the rank intervals in $I(v)$ with the purpose to include at most $O(\log n)$ intersection points for each $a\in C_A(v)$. 

\begin{figure}[htbp]
   \centering
   \subfigure[One intersection on both ends]{\includegraphics[page=1,scale=.75]{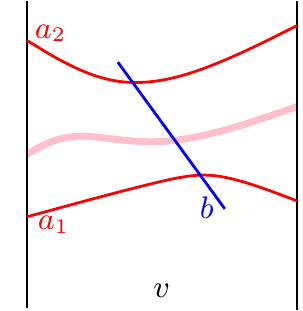}\quad
   \label{fig:xing1}} \hfill
   \subfigure[One intersection on one end and two on the other]{\quad\includegraphics[page=2,scale=.75]{cases} \quad
   \label{fig:xing12}} \hfill
   \subfigure[Two intersections on both ends -- nested]{\quad\includegraphics[page=3,scale=.75]{cases}\quad
   \label{fig:xing2n}} \hfill
   \subfigure[Two intersections on both ends -- disjoint]{\quad\includegraphics[page=4,scale=.75]{cases}
   \label{fig:xing2d}}
   \caption{Possible configurations of the intersection points of $b$ with the first and last red segments of its rank interval.}
   \label{fig:xing}
\end{figure}

The purpose of the next step is to avoid multiple intersections of a blue segment with a red one in its rank interval. Observe that there are three possibilities for the number of intersection points of $b$ with its lowest and highest intersected red segments $a_1$ and $a_2$: (1) Both red segments are intersected once (Fig.~\ref{fig:xing1}). In this case every red segment within the rank interval is also intersected exactly once. (2) One red segment is intersected once, and the other -- twice (Fig.~\ref{fig:xing12}). Let $a_1$ have one intersection point and $a_2$ two. Then all red segments intersected by $b$ once are the consecutive red segments following $a_1$,  all red segments with two intersections are consecutive segments preceding $a_2$, and all twice intersected segments yield nested intervals on $b$. Thus, when we split $b$ as in Step 3.1(c), one of the new segments has the same rank interval as $b$ and for the other we need to determine the new lower end of the interval (the upper end is $a_2$). (3) Both red segments are intersected twice. Then the $x$-intervals of their intersection points are either nested (Fig.~\ref{fig:xing2n}) or disjoint (Fig.~\ref{fig:xing2d}). If the intervals are nested, then all red segments of the rank interval are intersected twice in same manner, and both new subsegments of $b$ have the same rank interval. Otherwise, all red segments of the rank interval have to intersect the middle subsegment of $b$ but not necessarily the end subsegments. 
\begin{asubstep} 
	 For each node $v$ in $T$ and each $b\in E_B(v)$ do in parallel:
	 \begin{enumerate}
	    \item Find the rank interval of $C_A(v)$ intersected by $b$. 
      \item Compute the intersection points of $b$ with the lowest and with the highest intersected 
	 segment in $C_A(v)$. 
      \item If one interval end yields one and the other two intersection points (Fig.~\ref{fig:xing12}), split $b$ between the intersection points of the latter rank and find the new rank interval for one of the resulting segments (see below).
      \comment{
      \item If both interval ends yield two intersection points:
	 \begin{enumerate}
	    \item If the intersection intervals are nested (Fig.~\ref{fig:xing2n}), split $b$ between the innermost intersection points, both resulting segments have the same rank intervals as $b$.
	    \item If the intersection intervals are disjoint (Fig.~\ref{fig:xing2d}), split $b$ between each pair of intersection points. The middle segment has the rank interval of $b$ the intervals of the two end segments have to be determined.
	       }
      \item If both interval ends yield two intersection points and the intersection intervals are:
	 \begin{enumerate}
	    \item nested (Fig.~\ref{fig:xing2n}) -- split $b$ between the innermost intersection points, both resulting segments have the same rank intervals as $b$.
	    \item disjoint (Fig.~\ref{fig:xing2d}) -- split $b$ between each pair of intersection points. The middle segment has the rank interval of $b$ the intervals of the two end segments have to be determined.
	 \end{enumerate}
      \item Record the resulting interval(s) in the interval set $I(v)$.
	 \end{enumerate}
\end{asubstep}

Now we can be sure that each blue segment intersects each red segment in $C_A(v)$ at most once within the slab of $v$.

\begin{asubstep}
 For each node $v\in T$ in parallel construct an interval tree $T_I(v)$ for the set of rank intervals~$I(v)$. 
\end{asubstep}

\begin{figure}[htbp]
   \centering
      \includegraphics[page=1,scale=.75]{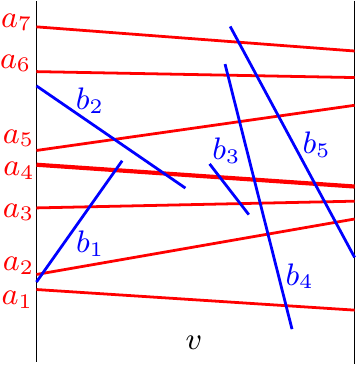} \hfill \includegraphics[page=2,scale=.75]{intervals} \hfill
      {\footnotesize
      \begin{tabular}[b]{lcccc}
	 & rank &prefix-  &prefix-& new \\
	 &  interval & min & max & interval(s)\\
	 $b_1$ & $[2,4]$ & 2 & 4 & $[2,4]$ \\
	 $b_2$ & $[4,5]$ & 2 & 5 & $[5,5]$ \\
	 $b_3$ & $[3,4]$ & 2 & 5 & empty \\
	 $b_4$ & $[1,6]$ & 1 & 6 & $[1,1],[6,6]$ \\
	 $b_5$ & $[2,7]$ & 1 & 7 & $[7,7]$ \\ 
      \end{tabular}}
   \caption{Segments in $C_A(v)$ and $E_B(v)$ corresponding to one node of $T_I(v)$ with the reference rank~$4$. In the right picture the removed parts of the intervals are denoted by dashed lines.}
   \label{fig:int}
\end{figure}

The blue segments whose rank intervals are stored in the same node $u$ of $T_I(v)$ all intersect the same red segment $a\in C_{A}(v)$ -- the segment with the reference rank of $u$, hereafter called the \emph{reference segment} of $u$. Thus, we can order them by the $x$-coordinate of their  intersection with $a$.  
If two blue segments $b_1, b_2$ intersect some red segments the order of the intersection points will be the same on all of these red segments. Therefore, if  $b_1$'s intersection with $a$ has a lower $x$-coordinate than that of $b_2$, we can remove from the rank interval of $b_2$ those elements that are in the interval of $b_1$ without losing significant intersection points.
\begin{asubstep}
	 For each node $v$ in $T$ and each node $u$ in $T_I(v)$ in parallel do:
	 \begin{enumerate} 
	    \item Sort the rank intervals of $u$ by the $x$-coordinate of the intersection with the reference segment  of $u$.
	    \item For the resulting ordered sequence compute the prefix-maxima of the highest ranks and the prefix-minima of the lowest ranks.
	    \item Replace the $i$-th interval $[r_{i,1},r_{i,2}]$, $i=2,3,\dots$, by two intervals\\ $\left[ r_{i,1},\min_{i-1}-1 \right]$ and $\left[ \max_{i-1}+1,r_{i,2} \right]$, where $\min_{i-1}=\min_{k=1,\dots i-1}r_{k,1}$ and  $\max_{i-1}=\max_{k=1,\dots i-1} r_{k,2}$ (see Fig.~\ref{fig:int}).
	    \item Discard all empty intervals, i.e., intervals $[r_1,r_2]$ where $r_2<r_1$.
	 \end{enumerate}
\end{asubstep}

The prefix-minimum ($\min_i$) and the prefix-maximum ($\max_i$) for the $i$-th interval in the ordered sequence  from Step 3.3 give us the range of ranks of the red segments that are already intersected by at least one of the corresponding blue segments $b_1,\dots,b_i$ before they (possibly) encounter an intersection with the segment $b_{i+1}$. Thus, we can remove the $[\min_i,\max_i]$ interval from the interval of $b_{i+1}$. 

After the interval spliting and shrinking $T_I(v)$ is no longer an interval tree, but the remaining intervals have the following properties:
The rank intervals  stored in one node of $T_I(v)$ are now disjoint, i.e., the rank of every segment $a\in C_A(v)$ is contained in at most one interval of a single node. The number of intervals in $T_I(v)$ is at most twice  the original number. Since every rank is stored in at most one node of each level of $T_I(v)$, the rank of every $a\in C_A(v)$ is contained in at most $O(\log n)$ intervals in $T_I(v)$. An example of such interval reduction for one node of the interval tree $T_I(v)$ is given in Figure~\ref{fig:int}.

Finally, we can reorganize $T_I(v)$ and compute the intersection points:

\begin{asubstep}
	 For each node $v\in T$ in parallel rebuild the interval tree $T_I(v)$ for the new  intervals.
\end{asubstep}
\begin{asubstep}
	 For each red segment $a\in C_A(v)$ in parallel find all $b\in E_B(v)$ whose  intervals in $T_I(v)$ contain the rank of $a$. Record the intersection point with the minimal $x$-coordinate. 
\end{asubstep}

The last step is to select the intersection points to report from the candidate intersections:
\begin{astep}[1]
   Sort the candidate intersection points computed in Steps 2 and 3 lexicographically by ($a_i$,$x$-coordinate), where $a_i$ is an identifier of the red segment in $A$.
   For each $a\in A$ report the intersection point with the minimal $x$-coordinate.
\end{astep}

In the next section we prove that the described algorithm correctly reports the first intersections of each segment in $A$ and do the counting to show the workload claimed in Theorem~\ref{la_2}.

\section{Proof of Theorem \ref{la_2}}

The correctness of the algorithm is based on the fact that in Steps 2 and 3 of the algorithm we consider all types of possible intersections and for each segment $a\in A$ discard only such intersections of each type which cannot have the minimal $x$-coordinate. We elaborate on the latter claim:

Consider the intersections of type (1) and (2), i.e., intersections within the vertical slab of a node $v$ of $T$ of a segment $a$ in $C_A(v)$ or in $E_A(v)$ with the segments in $C_B(v)$, which are handled in Step 2 of the algorithm. Since there are no blue-blue intersections, the segments of $C_B(v)$ partition the vertical slab corresponding to the $x$-interval of $v$ into (curvilinear) quadrilaterals. By locating the  neighbors of $a$ in $C_B(v)$ at the leftmost point of $a$ within the vertical slab of $v$ we find the quadrilateral in which $a$ ``enters'' the slab of $v$. Clearly, the first intersection for $a$ within the slab of $v$  (if any exists) is with one of its  blue neighbors. Because all curve segments are $x$-monotone, this is also the intersection with the lowest $x$-coordinate within the interval of $v$. We can safely discard all further intersections of $a$ with $C_B(v)$.

In Step 3 we initially capture all intersections of type (3) in the rank intervals without explicitly computing each intersection. 
Since our curve segments are well-behaved and we split the blue segments into subsegments that intersect each red curve at most once, if two blue sub-segments  intersect both two red segments $a_1$ and $a_2$, then the $x$-order of their intersection points is the same for $a_1$ and $a_2$. Thus, when we exclude the rank of a red segment $a$ from the interval of a blue sub-segment $b$ in Step 3.3, we know that there is another blue sub-segment,  whose intersection with $a$ has a lower $x$-coordinate. The intersection of $a$ and $b$ can then safely be discarded.

Next we prove the running time and the resource usage of the algorithm claimed in Theorem~\ref{la_2}:
Step~1 can be performed in $O(\log n)$ time on  $O(n)$ processor CREW-PRAM, see~\cite{Aggarwal:1988sf}. A segment tree uses $O(n\log n)$ space. 

In Step~2 we assign to each segment $a\in A$ a processor which traverses the segment tree $T$ and for each node $v$, with $a\in C_A(v)\cup E_A(v)$ performs a binary search on $C_B(v)$ to determine the  neighbors of $a$. The intersection computation is performed in constant time. This gives us $O(\log n)$ time for each of the $O(\log n)$ levels of $T$, i.e., $O(\log^2 n)$ time with $O(n)$ processors for all candidate intersections of type (1) and (2). Alternatively, we could assign a processor to each occurrence of a segment $a$ in cover- and end-lists of $T$ and find all candidates of type (1) and (2) in $O(\log n)$ time using $O(n\log n)$ processors. The first variant gives us $O(n)$ candidate intersections in Step 2, the second -- $O(n\log n)$. 

In Step 3.1 using one processor for each $b\in B$ we can find in time $O(\log^2 n)$ all rank intervals for the intersections of type (3) in $T$. Since each $b$ occurs in the end lists of at most two nodes per level of the segment tree $T$, each $b$ produces $O(\log n)$ such intervals. Sorting all $O(n\log n)$ boundaries of the rank intervals lexicographically by $(v,r)$, where $v$ is the node of $T$ that produced the interval, and $r$ is a boundary of the interval, we get the grouping of the intervals by the nodes of $T$ and sorting the interval boundaries in one step. This sorting can be performed in $O(\log^2 n)$ time using $O(n)$ processors.
Within the same time and processor bounds we can construct all interval trees in parallel (Step 3.2). Since an interval tree uses linear storage, the total space requirements remain $O(n\log n)$.

The operations in Step 3.3 involve sorting lists with a total number of $O(n\log n)$ elements, and performing prefix-max and prefix-min computations on these lists. These operations stay within the stated resource bounds. Splitting of the intervals and eventually deleting some of them can be performed independently for each interval, i.e., in $O(\log n)$ time by $O(n)$ processors in total. The number of the intervals is at most doubled in Step 3.3, thus the storage requirement remains $O(n\log n)$ and Step 3.4 has the same resource requirements as Step 3.2.

In Step 3.5 using one processor for each red segment $a$ we can traverse the segment tree $T$ and in each node $v$, such that $a\in C_A(v)$ find all intervals in $T_I(v)$ that contain the rank of $a$ in $C_A(v)$. As mentioned above, there are at most $O(\log n)$ such intervals in $T_I(v)$. Each interval corresponds to a blue segment in $E_B(v)$. Instead of reporting all intervals we compute the intersection of $a$ with the corresponding blue segment and keep track of the intersection point with minimal $x$-coordinate. The computation of all intersection points of $a$ stored in $T_I(v)$ is performed in $O(\log n)$ time, which gives us total of $O(\log^2 n)$ time over all levels of $T$. 
The number of candidate intersections produced in Step~3 is $O(n)$ -- at most one for each red segment.

Alternatively, all operations of Step~3 can be performed in $O(\log n)$ time on $O(n\log n)$ processors.

In Step 4 we sort $O(n)$ candidate intersection points by $(a,x)$, where $a$ is the number of the red segment and $x$ is the  $x$-coordinate of the intersection point, and take the minimum for each red segment. 
Step 4 can be performed in $O(\log n)$ time using $O(n)$ processors.
Which concludes the proof of Theorem~\ref{la_2}.

Throughout the algorithm we clearly use concurrent read operations, but for each write operation for each processor we can independently  determine a unique  storage slot. Thus the algorithm is designed for the CREW-PRAM model.

\section{Concluding Remarks}

The algorithm presented here can further be accelerated to perform Step 2 in $O(\log n)$ time using $O(n)$ processors by applying the fractional cascading technique \cite{Chazelle86fractionalcascading:}, \cite{68482}. This technique simplifies the iterative search for a key in multiple ordered lists and allows to perform the search of a segment $a$ in all $C_B$-lists of $T$ in $O(\log n)$ time, i.e., taking $O(\log n)$ time on $O(n)$ processors for Step 2. Steps 1 and 4 have already these time and processor bounds. But it is not clear whether the performance of Step 3 can be improved.
%
%

A further interesting related problem is matching geometric shapes under transformations (e.g., translations, rotations, scalings) with respect to the Hausdorff distance, i.e., find a transformation of one of the shapes such that the Hausdorff distance is minimized, for sequential algorithms see \cite{alt:behrends}.

Another interesting distance measure is the Frech\'et  distance \cite{ag-cfdbt-95}. A parallel solution to the decision problem, i.e., testing whether the Frech\'et distance between two polygonal curves is at most some given value $\varepsilon$, should be uncomplicated, since the sequential algorithm uses divide-and-conquer technique. We are currently working on a parallel algorithm for the computation problem, for which the sequential algorithm uses parametric search. 

\bibliographystyle{abbrv}
\bibliography{../bibl}

\begin{thebibliography}{10}

\bibitem{Aggarwal:1988sf}
A.~Aggarwal, B.~Chazelle, L.~Guibas, C.~{\'O}'D{\'u}nlaing, and C.~Yap.
\newblock Parallel computational geometry.
\newblock {\em Algorithmica}, 3(1):293--327, 1988.

\bibitem{alt:behrends}
H.~Alt, B.~Behrends, and J.~Bl{\"o}mer.
\newblock Approximate matching of polygonal shapes.
\newblock {\em Annals of Mathematics and Artificial Intelligence}, 13:251--265,
  1995.

\bibitem{ag-cfdbt-95}
H.~Alt and M.~Godau.
\newblock Computing the {F}r\'echet distance between two polygonal curves.
\newblock {\em Internat. J. Comput. Geom. Appl.}, 5:75--91, 1995.

\bibitem{as-chdco-08}
H.~Alt and L.~Scharf.
\newblock Computing the {H}ausdorff distance between curved objects.
\newblock {\em Int. J. Comput. Geometry Appl.}, 18(4):307--320, August 2008.

\bibitem{68482}
M.~J. Atallah, R.~Cole, and M.~T. Goodrich.
\newblock Cascading divide-and-conquer: a technique for designing parallel
  algorithms.
\newblock {\em SIAM J. Comput.}, 18(3):499--532, 1989.

\bibitem{Bentley:1980:OWC:1311064.1311126}
J.~L. Bentley and D.~Wood.
\newblock An optimal worst case algorithm for reporting intersections of
  rectangles.
\newblock {\em IEEE Trans. Comput.}, 29(7):571--577, July 1980.

\bibitem{808674}
B.~Chazelle.
\newblock Intersecting is easier than sorting.
\newblock In {\em STOC '84: Proceedings of the sixteenth annual ACM symposium
  on Theory of computing}, pages 125--134, New York, NY, USA, 1984. ACM.

\bibitem{Chazelle86fractionalcascading:}
B.~Chazelle and L.~J. Guibas.
\newblock Fractional cascading: I. a data structuring technique.
\newblock {\em Algorithmica}, 1:133--162, 1986.

\bibitem{4M}
M.~de~Berg, O.~Cheong, M.~van Kreveld, and M.~Overmars.
\newblock {\em Computational Geometry}.
\newblock Springer, Berlin, third edition, 2008.

\bibitem{Goodrich:1989:ILS:72935.72950}
M.~T. Goodrich.
\newblock Intersecting line segments in parallel with an output-sensitive
  number of processors.
\newblock In {\em Proceedings of the first annual ACM symposium on Parallel
  algorithms and architectures}, SPAA '89, pages 127--137, New York, NY, USA,
  1989. ACM.

\bibitem{DBLP:journals/algorithmica/GoodrichOY93}
M.~T. Goodrich, C.~{\'O}'D{\'u}nlaing, and C.-K. Yap.
\newblock Constructing the voronoi diagram of a set of line segments in
  parallel.
\newblock {\em Algorithmica}, 9(2):128--141, 1993.

\bibitem{Goodrich:1990:PMV:98524.98539}
M.~T. Goodrich, S.~B. Shauck, and S.~Guha.
\newblock Parallel methods for visibility and shortest path problems in simple
  polygons (preliminary version).
\newblock In {\em Proceedings of the sixth annual symposium on Computational
  geometry}, SCG '90, pages 73--82, New York, NY, USA, 1990. ACM.

\bibitem{Preparata:1985:CGI:4333}
F.~P. Preparata and M.~I. Shamos.
\newblock {\em Computational geometry: an introduction}.
\newblock Springer-Verlag New York, Inc., New York, NY, USA, 1985.

\bibitem{Tamassia:1989:OPA:72935.72978}
R.~Tamassia and J.~S. Vitter.
\newblock Optimal parallel algorithms for transitive closure and point location
  in planar structures.
\newblock In {\em Proceedings of the first annual ACM symposium on Parallel
  algorithms and architectures}, SPAA '89, pages 399--408, New York, NY, USA,
  1989. ACM.

\end{thebibliography}

\end{document}